\documentstyle[sprocl]{article}
\begin{document}

\begin{flushright}
IUHET 339\\
August 1996\\
\medskip
\end{flushright}

\title{CPT, STRINGS, AND NEUTRAL-MESON OSCILLATIONS}

\author{ V.\ ALAN KOSTELECK\'Y}

\address{Physics Department, Indiana University,\\ 
Bloomington, IN 47405, U.S.A.}

\maketitle\abstracts{
A mechanism for spontaneous CPT breaking appears in string theory.
Possible implications for present-energy 
particle models are discussed.
A realistic string theory might exhibit CPT violation at levels
detectable in current or future experiments.
Bounds on CPT from neutral-meson oscillations are considered.
}

\def\al{\alpha}
\def\be{\beta}
\def\ga{\gamma}
\def\de{\delta}
\def\ep{\epsilon}
\def\ve{\varepsilon}
\def\ze{\zeta}
\def\et{\eta}
\def\th{\theta}
\def\vt{\vartheta}
\def\io{\iota}
\def\ka{\kappa}
\def\la{\lambda}
\def\vpi{\varpi}
\def\rh{\rho}
\def\vr{\varrho}
\def\si{\sigma}
\def\vs{\varsigma}
\def\ta{\tau}
\def\up{\upsilon}
\def\ph{\phi}
\def\vp{\varphi}
\def\ch{\chi}
\def\ps{\psi}
\def\om{\omega}
\def\Ga{\Gamma}
\def\De{\Delta}
\def\Th{\Theta}
\def\La{\Lambda}
\def\Si{\Sigma}
\def\Up{\Upsilon}
\def\Ph{\Phi}
\def\Ps{\Psi}
\def\Om{\Omega}
\def\mn{{\mu\nu}}
\def\cl{{\cal L}}
\def\fr#1#2{{{#1} \over {#2}}}
\def\prt{\partial}
\def\ap{\al^\prime}
\def\apt{\al^{\prime 2}}
\def\apth{\al^{\prime 3}}
\def\pt#1{\phantom{#1}}
\def\vev#1{\langle {#1}\rangle}
\def\bra#1{\langle{#1}|}
\def\ket#1{|{#1}\rangle}
\def\bracket#1#2{\langle{#1}|{#2}\rangle}
\def\expect#1{\langle{#1}\rangle}
\def\sbra#1#2{\,{}_{{}_{#1}}\langle{#2}|}
\def\sket#1#2{|{#1}\rangle_{{}_{#2}}\,}
\def\sbracket#1#2#3#4{\,{}_{{}_{#1}}
 \langle{#2}|{#3}\rangle_{{}_{#4}}\,}
\def\sexpect#1#2#3{\,{}_{{}_{#1}}\langle{#2}\rangle_{{}_{#3}}\,}
\def\half{{\textstyle{1\over 2}}}
\def\frac#1#2{{\textstyle{{#1}\over {#2}}}}
\def\ni{\noindent}
\def\lsim{\mathrel{\rlap{\lower4pt\hbox{\hskip1pt$\sim$}}
    \raise1pt\hbox{$<$}}}
\def\gsim{\mathrel{\rlap{\lower4pt\hbox{\hskip1pt$\sim$}}
    \raise1pt\hbox{$>$}}}
\def\sqr#1#2{{\vcenter{\vbox{\hrule height.#2pt
         \hbox{\vrule width.#2pt height#1pt \kern#1pt
         \vrule width.#2pt}
         \hrule height.#2pt}}}}
\def\square{\mathchoice\sqr66\sqr66\sqr{2.1}3\sqr{1.5}3}

\def\Re{\hbox{Re}\,}
\def\Im{\hbox{Im}\,}

\newcommand{\beq}{\begin{equation}}
\newcommand{\eeq}{\end{equation}}
\newcommand{\bea}{\begin{eqnarray}}
\newcommand{\eea}{\end{eqnarray}}
\newcommand{\rf}[1]{(\ref{#1})}

Local Lorentz-invariant field theories of point particles
are known to exhibit invariance under the combined 
discrete-symmetry operations of charge conjugation C,
parity reversal P, and time reversal T.\cite{cpt}
This CPT theorem
has been experimentally tested in many systems,\cite{pdg}
including high-precision interferometric measurements 
using neutral kaons.
Modern searches for CPT violation 
can be viewed as limiting 
small effects that might arise from non-particle
physics at a level deeper than the 
standard model,
such as string theory.\cite{kp1}

The extended nature of strings
means that the usual derivation of the CPT theorem
does not hold in general.
This talk provides a short review of a mechanism
for spontaneous CPT breaking in string theory
and its possible observable consequences.
Details can be found in the original 
literature.\cite{kp1}$^-$\cite{kk}

It is known that Lorentz symmetry can be spontaneously
broken in string theory\cite{ks}
as a result of interactions that cannot appear in
a normal renormalizable gauge theory in four dimensions.
This can be accompanied by CPT violation.\cite{kp1}
The controlling interactions are consistent at the string level,
where there are infinitely many particle fields
due to the extended nature of the string.
When scalars condense in the vacuum,
the stringy interactions can trigger instabilities
in effective potentials for tensor fields.
The resulting vacuum values can spontaneously break
Lorentz and CPT invariance.

Some explicit evidence exists that these effects
occur in the open bosonic string.\cite{kp2}
The action of the string field theory can be  
calculated in a level-truncation scheme,
and the equations of motion can be solved 
for extrema of the action.
As expected,
the results include solutions with nonzero expectation
values for Lorentz tensors.

Assuming that the Universe is accurately described
at a fundamental level by a realistic string theory,
then it is possible that CPT-breaking contributions
could appear in the low-energy (compactified) limit.
Generic terms can be obtained that would arise
in a low-energy four-dimensional effective theory, 
which presumably would involve the standard model.
For example,
a contribution to the low-energy lagrangian could have 
the form:~\cite{kp2,kps}
\beq
\cl \supset \fr {\la} {M^k}
\vev{T}\cdot\overline{\ps}\Ga(i\prt )^k\ps
+ {\rm h.c.}
\quad .
\label{a}
\eeq
For neutral-meson oscillations,
$\ps$ can be regarded as a quark field in a meson,
coupled through a gamma-matrix structure $\Ga$
and possible derivative factors 
to a nonzero vacuum value $\vev{T}$ 
of a Lorentz tensor $T$.
The quantity $\la$ is a dimensionless constant,
while $M$ is a large mass scale
such as the Planck mass.

The absence of large CPT violation in nature
means that any effects of this type must be suppressed.
A natural dimensionless ratio in the theory
is $r = m_l/M$,
where $m_l$ is a low-energy scale.
Suppression by this small quantity 
could generate effects in neutral-meson systems
that are comparable in size to or smaller than
current and near-future experimental 
sensitivities.\cite{kp1,kps}

This possibility can be studied in more detail using 
an effective hamiltonian $\La$
controlling the time development
of the neutral-meson states.
In the standard approach,
CPT violation appears in $\La$ through a
phenomenological complex parameter $\de_P$,
with the subscript $P$ representing 
one of the four neutral mesons
$K$, $D$, $B_d$, and $B_s$.
In the context of the string scenario,
the expression \rf{a} provides a means
of relating $\de_P$ to more basic quantities.
It follows that\cite{kp2,kps}
\beq
\de_P = i \fr{h_{q_1} - h_{q_2}}
{\sqrt{\De m^2 + \De \ga ^2/4}} e^{i\hat\ph}
\quad ,
\label{b}
\eeq
where $\De m$ and $\De\ga$ are
the experimentally observable mass and rate differences
for the $P$ mesons,
with $\hat\ph = \tan^{-1}(2\De m/\De \ga)$,
and where 
$h_{q_j}=r_{q_j}\la_{q_j}\vev{T}$, $j = 1,2$,
are coefficients for the two valence quarks of $P$
that depend on quantities in \rf{a}
and on parameters $r_{q_j}$ arising 
from the quark-gluon sea.
Equation \rf{b} can be used to derive relationships
between the real and imaginary parts of the parameters $\de_P$:
\beq
\Im \de_P = \pm \cot\hat\ph ~\Re\de_P
\quad .
\label{c}
\eeq
These predictions of the string scenario for the $P$ mesons
are consequences of the hermiticity of the fundamental
string action.

One important feature that emerges from the above results
is that the size of any CPT violation arising in this
context is potentially very different in different 
meson systems. 
The presence of constant dimensionless couplings
in Eq.\ \rf{a} is reminiscent of the 
standard-model Yukawa couplings,
which are known to vary over many orders of magnitude.
It is therefore of interest to test for CPT breaking
in all neutral-meson systems,\cite{kp2}
not only the $K$ case.

Each neutral-meson system can be examined 
to determine the limits on CPT violation that could
be placed by present and near-future 
experiments using either correlated or uncorrelated
mesons.\cite{kps}$^{-}$\cite{kk}
Here is a short summary of a few results
emerging from these analyses.

$\bullet$
Currently, 
only the $K$-system CPT-violating parameter $\de_K$
has been bounded experimentally. 
The advent of high-statistics kaon and $\ph$ factories
suggests that improved limits,
perhaps even to one part in $10^{5}$,
are likely soon.

$\bullet$
Since mixing in the $D$ system is known to be no
larger than about 5\% and may be much smaller,
bounding CPT violation there is difficult.
At a $\ta$-charm factory,
however,
several significant experiments could be envisaged,
including ones limiting some kinds 
of $K$-system CPT violation
as well as direct and indirect 
CPT violation in the $D$ system.\cite{ck}

$\bullet$
Especially interesting limits 
are feasible for the neutral-$B_d$ system,
for which the heavy $b$ quark is involved
and CPT violation could be enhanced.
Indeed,
it is still possible that indirect CPT violation 
exceeds the more conventional indirect T violation 
in this system.
No CPT limit has yet been published,
but recent Monte-Carlo simulations with realistic 
data incorporating experimental backgrounds and
acceptances suggest that 
sufficient data have already been taken 
to place a bound on $\de_B$ around the 10\% level.\cite{kk}
Machines currently being developed would have the
capability for further improvements.

\medskip

I thank Don Colladay, Rob Potting, Stuart Samuel, 
and Rick Van Kooten for many discussions on the subject
of CPT violation.


\begin{thebibliography}{xx}
\baselineskip=11pt
\bibitem{cpt}
See, for example, R.F. Streater and A.S. Wightman,
\it PCT, Spin and Statistics, and All That, \rm
Benjamin Cummings, Reading, 1964.

\bibitem{pdg}
See, e.g., Review of Particle Properties,
Phys.~Rev.~D {\bf 54}, 1 (1996).

\bibitem{kp1}
V.A. Kosteleck\'y and R. Potting,
Nucl.~Phys.~B {\bf 359}, 545 (1991).

\bibitem{ks}
V.A. Kosteleck\'y and S. Samuel,
Phys.~Rev.~D {\bf 39}, 683 (1989);
\it ibid., \rm
{\bf 40}, 1886 (1989);
Phys.~Rev.~Lett. {\bf 63}, 224 (1989);
\it ibid., \rm
{\bf 66}, 1811 (1991).

\bibitem{kp2}
V.A.\ Kosteleck\'y and R.\ Potting,
Phys.\ Lett.\ B {\bf 381}, 89 (1996). 

\bibitem{kps}
V.A.~Kosteleck\'y and R.~Potting,
Phys.~Rev.~D {\bf 51}, 3923 (1995);
see also V.A.~Kosteleck\'y, R.~Potting and S.~Samuel,
in S.~Hegarty et al., eds.,
\it Proc.\ 1991 Joint Intl.\ Lepton-Photon Symp.\ 
and Europhys.\ Conf.\ on High Energy Phys.\ \rm
(World Scientific 1992);
V.A.~Kosteleck\'y and R.~Potting,
in D.B.~Cline, ed.,
\it Gamma Ray--Neutrino Cosmology and Planck Scale Physics \rm
(World Scientific 1993)
(hep-th/9211116).

\bibitem{ck}
D. Colladay and V.A. Kosteleck\'y,
Phys.~Rev.~D {\bf 52}, 6224 (1995);
these proceedings.

\bibitem{kk}
D. Colladay and V.A. Kosteleck\'y,
Phys.~Lett.~B {\bf 344}, 259 (1995);
V.A. Kosteleck\'y and R. Van Kooten,
Phys.~Rev.~D (1996), in press (hep-ph/9607449).

\end{thebibliography}
\end{document}